\documentstyle[12pt]{article}

\begin{document}

\vglue 1.0 cm

\begin{flushright}
  {TSU HEPI 2000-03 }   
\end{flushright}

\vglue 3.5cm

\begin{center}
{\Large \bf Logarithmic Potential Model of Quigg and Rosner as a Generalization
of Naive Quark Model }
\medskip

\vglue 1.0cm
{\bf A.\ A.\ Khelashvili\footnote {e-mail:temo@hepi.edu.ge}}
\medskip

{\it High Energy Physics Institute, Tbilisi State University, 380086, Tbilisi, 
Georgia}

\vglue 2.5cm

{\bf  { ABSTRACT }} \\
\vglue 2.0cm

\parbox{6.4in}{\leftskip=1.0pc

Exploiting the explicit mass formulae for logarithmic potential model of Quigg and Rosner it
 is shown that at least on the level of mass-relations this model reproduces the naive
 quark model relations and generalizes the last one in case of highly non-trivial
 potential. Generalization includes the relations for higher values of orbital quantum
 numbers. In particular, preditions for recently discovered atom-like P-states are no
 worse than for any other potential models.The advantage consists in simplicity of
 approach.  }

\end{center}

\newpage

\medskip

Examining the approximate equality of the spacing between the ground
states and first radial excitations in the Upsilon 
and $J/\Psi$ systems Quigg and Rosner \cite{1} demonstrated 
that the logarithmic potential is the only nonrelativistic 
potential in quantum mechanics that predicts
ind ependence of the radial excitations over the constituent masses.

They also calculated the numeric values for all observable physical
quantities for this potential. It turned out that from the phenomenological
point of view this potential is no worse then the other ones. So, though
the linear potential has more sound theoretical basis, the logarithmic one
also deserves attention.

In this note I will concentrate on some of the mass 
relations (sum rules) which can be derived explicitly 
only within the logarithmic potential model. In the literature 
it was never noticed that the logarithmic potential, beeing 
nonrelativistic, generalizes naive quark model. So its predictions 
are simple and can be formulated analytically. In the last years 
Martin's phenomenological potential \cite{2} $V(r)= - 0.8064+6.87
r^{0.1}$ has gained much interest. Coupled with the simplest 
spin-spin Hamiltonian this model leads to satisfactory results 
for lower levels of quarkonia.

Due to small exponent Martin's potential is quite close to the 
logarithmic one. So it's natural to expect that numerical results 
will be quite similar. And we will see below that in the latter 
case simple analytic results can be derived.

So, for each quark-antiquark pair we will use the following 
potential \cite{1}:
\begin{equation}
V_{ij}(r)=g_{ij}\ln\frac{r}{R_{ij}}
\label{1}
\end{equation}

Then it is easy to show that after introduction of the dimensionless 
parameter
$\rho=(2\mu_{ij}g_{ij})^{1/2}r$
(where $\mu_{ij}=m_im_j/(m_i+m_j)$ is a reduced mass) into the
Schr\"odinger equation,  masses of the states with some orbital
momentum $l$ and the radial quantum
number $n$ are given by \cite{1}\cite{3}:
\begin{equation}
M_n^l(q_i,\bar q_j)=m_i+m_j-
\frac{g_{ij}}{2}\ln(2\mu_{ij}R^2_{ij}g_{ij})+g_{ij} E_{nl}\ ,
\label{2}
\end{equation}
where the eigenvalues $ E_{nl}$
depend only on $n$ and $l$ and not on the other parameters --- namely,
the masses of constituents. Their numeric \cite{1} and also approximate analytical
expessions \cite{4} are
known but we will not need them now.

Of course one can take $g_{ij}$ and $R_{ij}$ flavour-dependent in order to fine-tune phenomenological 
predictions but hypothesis of flavour-independence, as in the Martin potential case, is more attractive. It was flavour 
independence that served as an argument for initial consideration of the Logarithmic Potential \cite{3} 
because of approximate equality of $\psi'-\psi$ and ${\cal Y}'-{\cal Y}$ splittings. It is clear from the very start that 
absolute flavour independence is not the case here, but existing difference can be attributed to the relativistic 
corrections.

In the case of flavour independence there arise number of mass relations. The one that interests us is the 
following:
\begin{equation}
M_{n_2}^{l_2}(Q\bar q)- M_{n_1}^{l_1}(Q\bar q)= M_{n_2}^{l_2}(Q\bar Q)- M_{n_1}^{l_1}
(Q\bar Q)\ .
\label{3}
\end{equation} 
This relation is symmetric under interchange of $Q$ and $q$ i.e. 
together with (\ref{3}) we have:
\begin{equation}
M_{n_2}^{l_2}(q\bar q)- M_{n_1}^{l_1}(q\bar q)= M_{n_2}^{l_2}(Q\bar Q)- M_{n_1}^{l_1}
(Q\bar Q)\ .
\label{4}
\end{equation} 
This relation is critical for estimation of flavour independence --- it gets less precise the lighter the quarks 
are. It is not surprising because the relativistic effects are more important for light quarks\footnote{As long as we 
neglect particle spins, and the mass relations are written in the center of mass of multiplets, we deal only with 
the spin-independent relativistic corrections.}. As for the relation
(\ref{3}), one can expect that the relativistic
corrections are of the same order for both sides: though in the
left-hand-side one of the quarks is light, but the
system is atom-like and so the light quark is predominantly located at
large "distances'' \cite{4} and it can lead
to effective damping of the matrix elements of the perturbation Hamiltonian.

The above considarations sound quite speculative until they are not
checked in actual calculations, but the
underlying physical picture seems quite realistic.

Formula (\ref{2}) leads to the rule for the level shifts in 
the different families
without making assumption of universality of constants \cite{3}:
\begin{equation}
\frac{
M_{n_1}^{l_1}(Q\bar Q)- M_{n_2}^{l_2}(Q\bar Q)
}
{
M_{{n'}_1}^{{l'}_1}(Q\bar Q)- M_{{n'}_2}^{{l'}_2}(Q\bar Q)}
=
\frac{
M_{n_1}^{l_1}(q\bar q)- M_{n_2}^{l_2}(q\bar q)
}
{
M_{{n'}_1}^{{l'}_1}(q\bar q)- M_{{n'}_2}^{{l'}_2}(q\bar q)
}
\label{5}
\end{equation}

These rules can be used to relate mass differences in the light and heavy
families.

Below we will consider some consequences for the examples where more or less
experimantal data is available for the states with $n=1$ and $l=0,1$

{\underline \bf 1.\ $(s\bar q)$ system ($q\equiv u$ or $d$):} \\
From (\ref{3}) we have:
\begin{equation}
M_1^1(s\bar q;1^+)= M_1^0(s\bar q;1^-)+[ M_1^1(s\bar s;1^{++})- M_1^0(s\bar s;1^{--})]\ .
\label{6}
\end{equation}
Using identification from \cite{6} we get:
\begin{equation}
M_1^1(s\bar q;1^+)=K^+(892)+[f_1(1530)-\Phi(1020)]\approx 1402MeV \ .
\label{7}
\end{equation}

Experimental data for the corresponding multiplet is \cite{6}: $2^+\ K^*_2(1430),$ $1^+\ K^*_1(1400),$ $ 
0^+\ K^*_0(1350)$. Hence $M_{cm}=(5 K^*_2+3 K^*_1+ K^*_0)/9\approx 1411MeV$.

{\underline \bf 2.\ $(c\bar s)$ system:} \\
In this case in correspondence with (\ref{3}) we have:
\begin{equation}
M^1_1(c\bar s;1^+)=D_s^*(2110)+[\chi_c(3522)-{}^J/{}_\psi(3097)]\approx 2535MeV\ .
\label{8}
\end{equation}
It is in good agreement with experimental value 2536MeV \cite{8}, and also with the value calculated by 
Martin \cite{2}, $2532$Mev.

{\underline \bf 3.\ $(c\bar q)$ system:} \\
The prediction is:
\begin{equation}
M^1_1(c\bar q;1^+)=D^*(2010)+[\chi_c(3522)-{}^J/{}_\psi(3097)]\approx 2435MeV\ .
\label{9}
\end{equation}
Candidates for the $1^+$ and $2^+$ states were found at 2424 and 2459MeV \cite{9}, respectively. 
Our result is in good agreement with them.

{\underline \bf 4.\ $(b\bar q)$ and $(b\bar s)$ systems:} \\
Corresponding formulae have the form:
$$
M^1_1(b\bar q;1^+)=B^*(5324.6\ ?)+[\chi_b(9900)-{\cal Y}(9460)]\approx 5764MeV\ .
$$
\begin{equation}
M^1_1(b\bar s;1^+)=B_s^*(\ 5416\ )+[\chi_b(9900)-{\cal Y}(9460)] \approx 5856MeV\  .
\label{10}
\end{equation}

Unfortunattely there is no valid experimental data yet to check these results. 
Readers well acquainted
with the spectroscopy of particles can find other examples of 
employing (\ref{3}) too.

What can be said in conclusion about the mass relation (\ref{3})?
It is designed in the way to comply with the naive
quark counting, being at the same time valid for far from trivial
type of potential (\ref{1}). It is worth noting here
that similar mass relations emerge also for the multiquark 
bound states and lead to fair predictions for the 
Baryon masses \cite{10},\cite{11}.

For multiquark systems it is very convenient to use the method of
hyperspheric harmonics \cite{10}. In the so-called diagonal
approximation this formalism for the logarithmic potential between
pairs  leads to the following mass formula for N-particle bound
state:
\begin{equation}
M(1,2,\ldots ,N)\equiv M_n^{\lambda(L)}=\sum_{i=1}^N m_i+
C_{[L][L]}-\frac{1}{2}\sum_{i<j}g_{ij}\ln\left[
\frac{2m_im_j}{m_i+m_j}R_{ij}^2(\sum_{k<l}g_{kl})
\right]
\label{11}
\end{equation}
$$\qquad +(\sum_{k<l}g_{kl})E_{n\lambda}\ .
$$
Here $\lambda=L+3(N-2)/2$ and $L$ is the grand momentum.
The usual QCD relation between the quarks' and
quark-antiquark coupling constants $g_{ij}\to \frac{1}{2}g_{ij}$
was also exploited in (\ref{11}).

The eigenvalues $E_{n\lambda}$ are also mass and coupling $g_{ij}$
independent. If the constituents have different masses then the
coefficients  $C_{[L][L]}$ depend on the masses and so the mass-dependence
arises in the orbital (but not radial) excitations.

We do not have mass-dependence in the $L=0$ case when:
$$
C_{[L=0][L=0]}=\frac{1}{2}(\sum_{k<l}g_{kl})
\left\{\Psi(\frac{3}{2})-\Psi(\frac{3(N-1)}{2})\right\}
$$
where $\Psi(z)$ is the logarithmic derivative of the Euler's
$\Gamma$-function.

Using these formulae we can obtain relations between spin 
averaged masses.
They have the naive quark counting form. E.g.\ \cite{10}:
$$
M(s\bar s) - M(c\bar c)=2[M(ssc)-M(ccs)]=\frac{2}{3}[M(sss)-M(ccc)]\ ,
$$
$$
M(c\bar s) - M(s\bar s)=M(ssc)-M(sss)\ ,
$$
$$
M(c\bar s) - M(c\bar c)=M(scc)-M(ccc)\ \mbox{\it etc.}
$$
These are the meson-baryon mass formulae.

One more relation for light quarks is the following \cite{11}:
$$
M(\bar qc) - M(\bar qs)=M(qqc)-M(qqs)\ .
$$
If we take into account also the spin interactions we'll get 
sum rules which fit with experiments quite well:
$$
M_{\Lambda_c}=M_\Lambda+\frac{1}{4}[M_D-M_K+3(M_{D^*}-M_{K^*})]\sim
{\mbox 2295Mev (exp.\ 2285Mev)}
$$
$$
M_{\Sigma_c}=M_\Sigma+\frac{1}{12}[7(M_D-M_K)+5(M_{D^*}-M_{K^*})]\sim
{\mbox 2448Mev (exp.\ 2455Mev)}
$$

So we see that the logarithmic potential model is a good
generalization of the naive quark model and is quite simple in use.

\end{document}